\documentclass[journal]{IEEEtran}

\usepackage{cite}
\usepackage{hyperref}

\usepackage{graphicx}
\usepackage{multirow}
\usepackage[normalem]{ulem}
\usepackage{color}
\usepackage{caption}

\usepackage{amsmath}
\usepackage{siunitx}

\usepackage{xr,soul} 
\usepackage[dvipsnames]{xcolor}
\usepackage[switch]{lineno}

\usepackage{acronym}
\acrodef{AER}[AER]{Address Event Representation}
\acrodef{FPGA}[FPGA]{Field Programmable Gate Array}
\acrodef{PFM}[PFM]{Pulse--Frequency Modulation}
\acrodef{PWM}[PWM]{Pulse--Width Modulation}
\acrodef{PID}[PID]{Proportional Integral Derivative}
\acrodef{PI}[PI]{Proportional Integral}
\acrodef{P}[P]{Proportional}
\acrodef{SNN}[SNN]{Spiking Neural Network}
\acrodef{WTA}[WTA]{Winner-Take-All}
\acrodef{YARP}[YARP]{Yet Another Robot Platform}
\acrodef{DYNAP-SE}[DYNAP-SE]{Dynamic Neuromorphic Asynchronous Processors}
\acrodef{SoC}[SoC]{System on chip}
\acrodef{CPU}[CPU]{central processing unit}
\acrodef{GUI}[GUI]{Graphical User Interface}
\acrodef{DTP}[DTP]{discrete target pursuit}
\acrodef{CoM}[CoM]{center of mass}
\acrodef{RMSE}[RMSE]{Average Root Mean Square Error}

\begin{document}
\title{Closed-loop spiking control on a neuromorphic processor implemented on the iCub}
%
%
%

\author{Jingyue Zhao,
        Nicoletta Risi, Marco Monforte, Chiara Bartolozzi,~\IEEEmembership{Member,~IEEE},
        Giacomo~Indiveri,~\IEEEmembership{Senior~Member,~IEEE}, and Elisa~Donati,~\IEEEmembership{Member,~IEEE}
\thanks{J. Zhao, N. Risi, E. Donati and G. Indiveri are with the Institute of Neuroinformatics, University of Zurich,  8006 Zurich Switzerland, and also with the ETH Zurich, 8092, Zurich, Switzerland (e-mail: {jzhao,nicoletta,elisa,giacomo}@ini.uzh.ch).}
\thanks{M. Monforte, C. Bartolozzi are with the Istituto Italiano di Tecnologia, Genova, Italy (email:{marco.monforte,chiara.bartolozzi}@iit.it)}
\thanks{This work is supported by the European Union's Horizon 2020 Marie Skłodowska-Curie Action grant NEPSpiNN (Grant No. 753470), the Forschungskredit grant FK-18-103, and the H2020 ERC project NeuroAgents (Grant No. 724295) }
}
\maketitle

\begin{abstract}

Despite neuromorphic engineering promises the deployment of low latency, adaptive and low power systems that can lead to the design of truly autonomous artificial agents, the development of a fully neuromorphic artificial agent is still missing. While neuromorphic sensing and perception, as well as decision-making systems, are now mature, the control and actuation part is lagging behind. 
In this paper, we present a closed-loop motor controller implemented on mixed-signal analog-digital neuromorphic hardware using a spiking neural network. The network performs a proportional control action by encoding target, feedback, and error signals using a spiking relational network. It continuously calculates the error through a connectivity pattern, which relates the three variables by means of feed-forward connections. Recurrent connections within each population are used to speed up the convergence, decrease the effect of mismatch and improve selectivity.
The neuromorphic motor controller is interfaced with the iCub robot simulator. We tested our spiking P controller in a single joint control task, specifically for the robot head yaw. The spiking controller sends the target positions, reads the motor state from its encoder, and sends back the motor commands to the joint. The performance of the spiking controller is tested in a step response experiment and in a target pursuit task. In this work, we optimize the network structure to make it more robust to noisy inputs and device mismatch, which leads to better control performances. 
\end{abstract}


%
\IEEEpeerreviewmaketitle

\section{Introduction}
\label{sec:intro}

\IEEEPARstart{N}{euromorphic} computation is a promising framework for the development of embedded and efficient systems that will allow artificial devices to interact with the real world in real-time, by exploiting computational principles derived from biological structures. The emulation of neurons and synapses dynamics in a compact and energy-efficient technology~\cite{chicca2014neuromorphic} makes \ac{SNN} suitable for embedded low-power applications such as autonomous robotics, prosthetics, and always-on wearable biomedical signals processing. In addition, the spiking nature of the silicon neurons is a natural match for robotic systems that need to be interfaced with biological signals.

Neuromorphic sensors and processors are being increasingly developed and integrated into robotic applications, especially where fast, compact and power-efficient devices are required~\cite{bartolozzi2011embedded, delbruck2013robotic}. Despite their extensive deployment, a fully closed-loop neuromorphic system is still missing. Motors are still being controlled with traditional approaches. 
Recent works~\cite{perez2013neuro, perez2014approach} presented fully spiking open-loop motor controllers implemented on \acp{FPGA}. Open-loop controllers, however, are known to be less accurate and reliable.
A simulated implementation of a closed-loop controller was proposed in~\cite{perez2017towards}, where the commands were sent to the motors as spike trains using \ac{PFM} rather than traditional \ac{PWM}. In such a way the motor commands are encoded in the signal frequency instead of the signal width, and the spikes can be sent directly to the motor without any conversion, decreasing latency and power consumption. The first implementation of \ac{PFM} control on a mixed-signal neuromorphic device was proposed in~\cite{donati2018open}, although the system was open-loop.

\begin{figure}[t]
    \centering
    \includegraphics[width=0.5\textwidth]{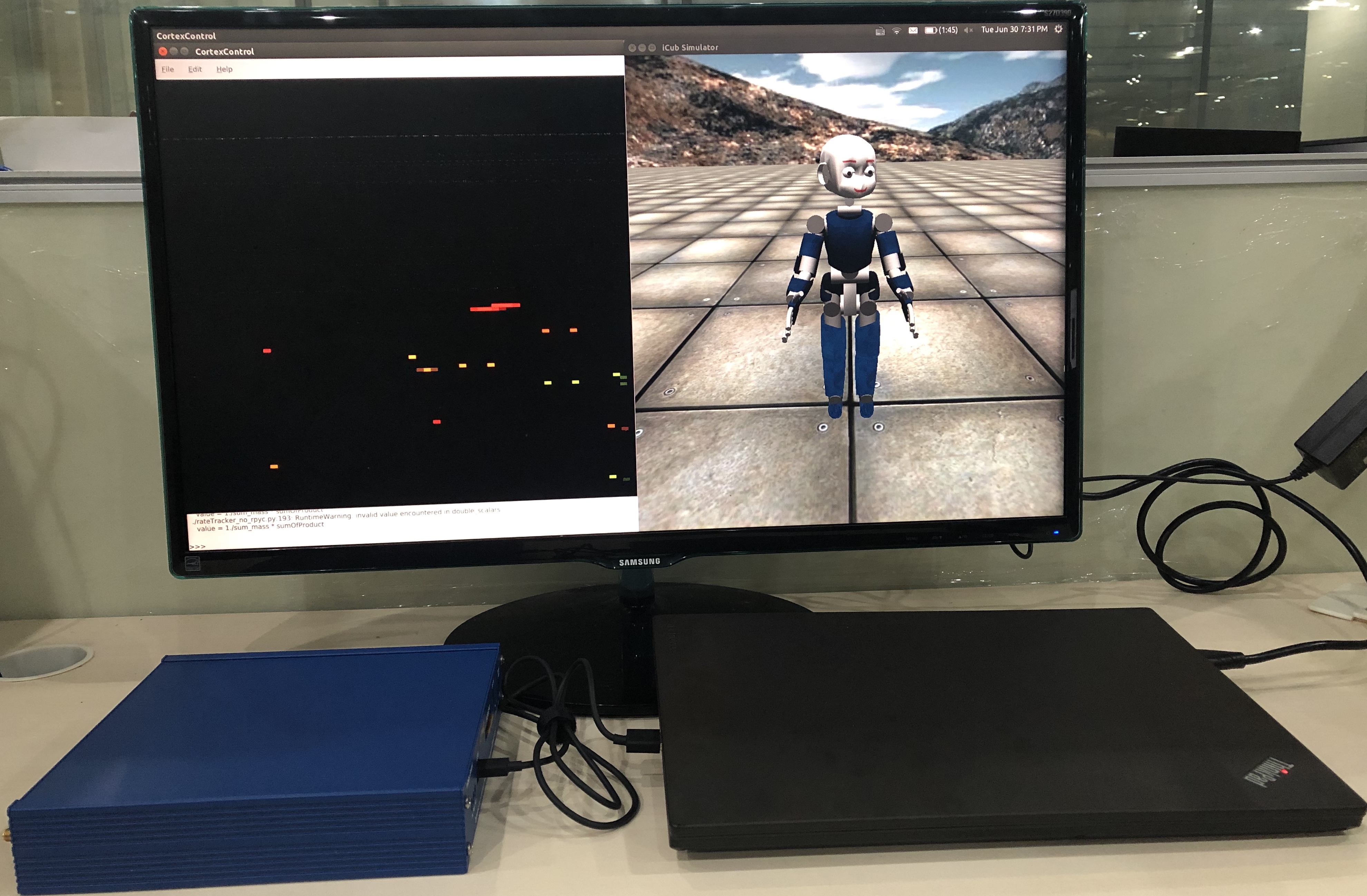}
    \caption{The setup used for the experiments, including the CTXCTL software, the iCub simulator (both running on the laptop) and the \ac{DYNAP-SE} (blue box on the left). The monitor shows the simulated iCub and (on the left) the spikes recorded from the \ac{DYNAP-SE}.}
    \label{fig:setup}
\end{figure}

A different approach consists in the use of \ac{SNN} to implement the traditional \ac{PID} controller. Different attempts have been made in this direction: on \acp{FPGA}~\cite{jimenez2012neuro} with the neuromorphic digital processor SpiNNaker~\cite{galluppi2014event}, and on a mixed analog-digital device~\cite{glatz2018adaptive}. In the latter, the \ac{SNN} is in charge of the error calculation and the motor command is encoded in the firing rate of the output population. The inherent mismatch problem in this type of encoding is solved by a preliminary calibration phase, where neurons with similar behavior are selected. However, this procedure is time-consuming and drastically reduces the number of available neurons to implement the \ac{SNN}. An alternative way to cope with mismatch is to resort to space coding, whereby the motor command is encoded by the identity of the neuron that is spiking, irrespective of its firing rate~\cite{zhao2020neuromorphic}. The spiking \ac{P} controller uses relational \ac{SNN}~\cite{diehl2016learning} to encode target, feedback and error. The latter is continuously computed through the three variables relation encoded in the customised feed-forward inter-population synapses.

This work extends the scope of~\cite{zhao2020neuromorphic} with the characterisation of the controller implemented on the \ac{DYNAP-SE} mixed-mode neuromorphic processor~\cite{moradi2017scalable}, controlling a single joint of the humanoid platform iCub. The system is characterised by its step response and evaluated in a target pursuit task.

As showed in the next pages, given the responsiveness and low error range of the control, the proposed architecture proves to be a promising spiking computational building block for a future more complex controller.

The overall system is modular, based on a robotic middleware and \ac{FPGA} components working with neuromorphic standard interfaces, this allows us to easily plug and play different neuromorphic platforms to benchmark the spiking controllers or perception modules implemented on different neuromorphic hardware.

\section{Materials and Methods}
\label{sec:mm}
To build a closed-loop neuromorphic motor controller, we interfaced a mixed analog-digital neuromorphic processor~\cite{moradi2017scalable} with a simulator of the humanoid robot iCub~\cite{tikhanoff2008open} to control the robot head yaw joint (\figurename~\ref{fig:yarp}).

\subsection{Neuromorphic hardware}
The mixed-signal neuromorphic chip used to implement the spiking motor control network is the \ac{DYNAP-SE}~\cite{moradi2017scalable}.  It integrates analog circuits that emulate the behavior of biological neurons and synapses, and digital logic circuits for configuration and communication based on the \ac{AER} protocol~\cite{deiss1999pulse}. 
The \ac{DYNAP-SE} processor consists of 4 chips, each composed of 1024 adaptive leaky integrate-and-fire neurons~\cite{indiveri2003low} that can receive inputs from 64 sources and send their output up to 4096 neurons. 
A \ac{SNN} can be implemented on the chip by setting neuron parameters and configuring their connections. 

\subsection{Robotic platform and system interfacing}
As test-bed for the characterisation of the proposed controller architecture, we resorted to the open-source iCub humanoid platform simulator iCubSim~\cite{tikhanoff2008open}, controllable using the \ac{YARP} middleware~\cite{metta2006yarp}. This configuration enabled the characterisation of the system using the robot by running simulations on a virtual platform and will allow us to easily deploy the system on the real iCub robot using the same interfaces. The \ac{YARP} Event-driven library~\cite{glover_etal18}, coupled with \ac{FPGA} and \ac{SoC} devices, enables the communication to different types of neuromorphic hardware and software computing modules. Supported neuromorphic hardware comprises vision sensors (ATIS~\cite{posch_etal} and DVS~\cite{lichtsteiner_etal08}), tactile sensors~\cite{mottoros_etal} and neuromorphic computing platforms, such as SpiNNaker~\cite{furber2014spinnaker}. In this work, we developed the supporting infrastructure and modules to integrate \ac{DYNAP-SE}. 
The physical system setup is shown in Fig.~\ref{fig:setup}. The configuration of the \ac{DYNAP-SE} chip is performed through a custom Python API (CTXCTL\footnote{\url{http://ai-ctx.gitlab.io/ctxctl/primer.html\#python-api}}). The communication between the iCub simulator and software modules running on the laptop and \ac{DYNAP-SE} are managed via an FPGA (Spartan-6 XC6SLX25).

\begin{figure}[t]
    \centering
    \includegraphics[width=0.5\textwidth]{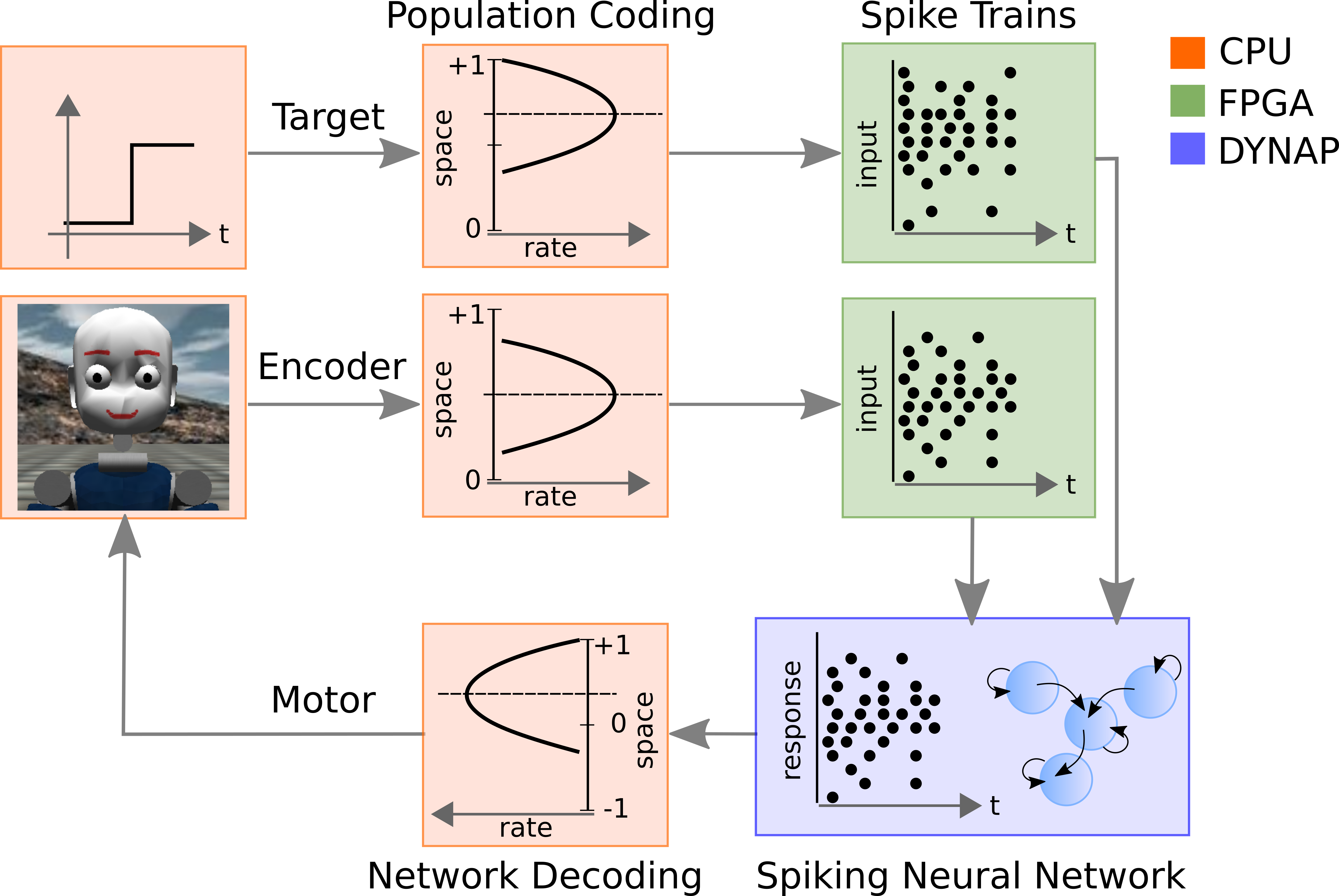}    \captionsetup{justification=centering}
    \caption{The overall pipeline, showing the implemented modules and the communication system between them.}
    \label{fig:yarp}
\end{figure}

Fig.~\ref{fig:yarp} shows the block diagram of the whole system. 
The target position and the current robot state (e.g. output of the encoders from the iCub simulator) are converted into space coding (on \ac{CPU}, orange blocks). The motor positions are encoded in an array, where the analog value of the target (or the motor state) is the center of a Gaussian distribution. The value of the Gaussian at each point represents the mean firing of a neuron sensitive to the target (or measured) motor position.
These analog values are sent to the FPGA (green blocks), where they are converted into Poisson spike trains and sent to the neuromorphic hardware using configurable spike generators.
The output spikes of the neural network implemented on the chip (blue block) are sent to the laptop via the \ac{FPGA}, and are decoded into a motor command that is sent to the iCub simulator.
The communication across all modules is supported by \ac{YARP}.


\subsection{Neuromorphic motor controller}
\begin{figure}[!t]
\centering
\includegraphics[width=0.48\textwidth]{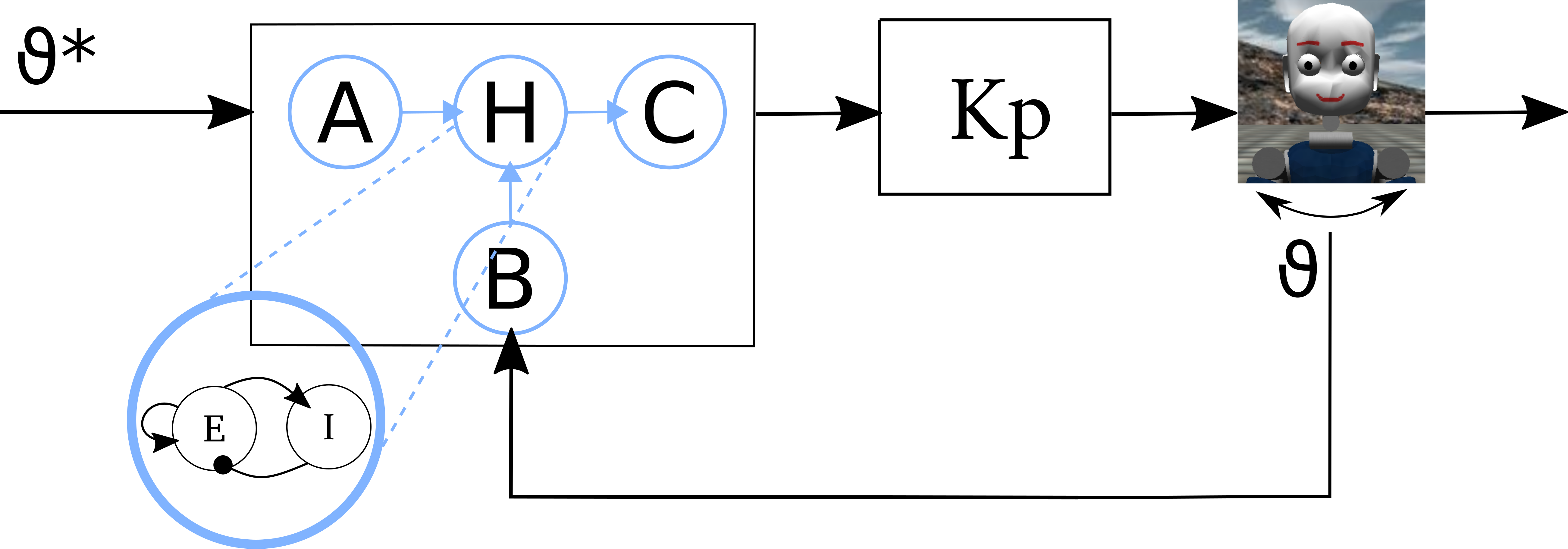}
\caption{Block diagram of the spiking \ac{P} controller. The threeway network output is multiplied by the proportional gain $K_p$ and sent to the iCub simulator.}
\label{fig:ctrl}
\end{figure}

\figurename~\ref{fig:ctrl} shows the block diagram of a simple \ac{P} controller. The desired (target) position of the motor $\theta^*$ is compared to the current position of the motor, measured by the encoder, $\theta$. The motor command is then updated proportionally to their difference (the position error).

\begin{figure*}[!t]
\centering
\includegraphics[width=\textwidth]{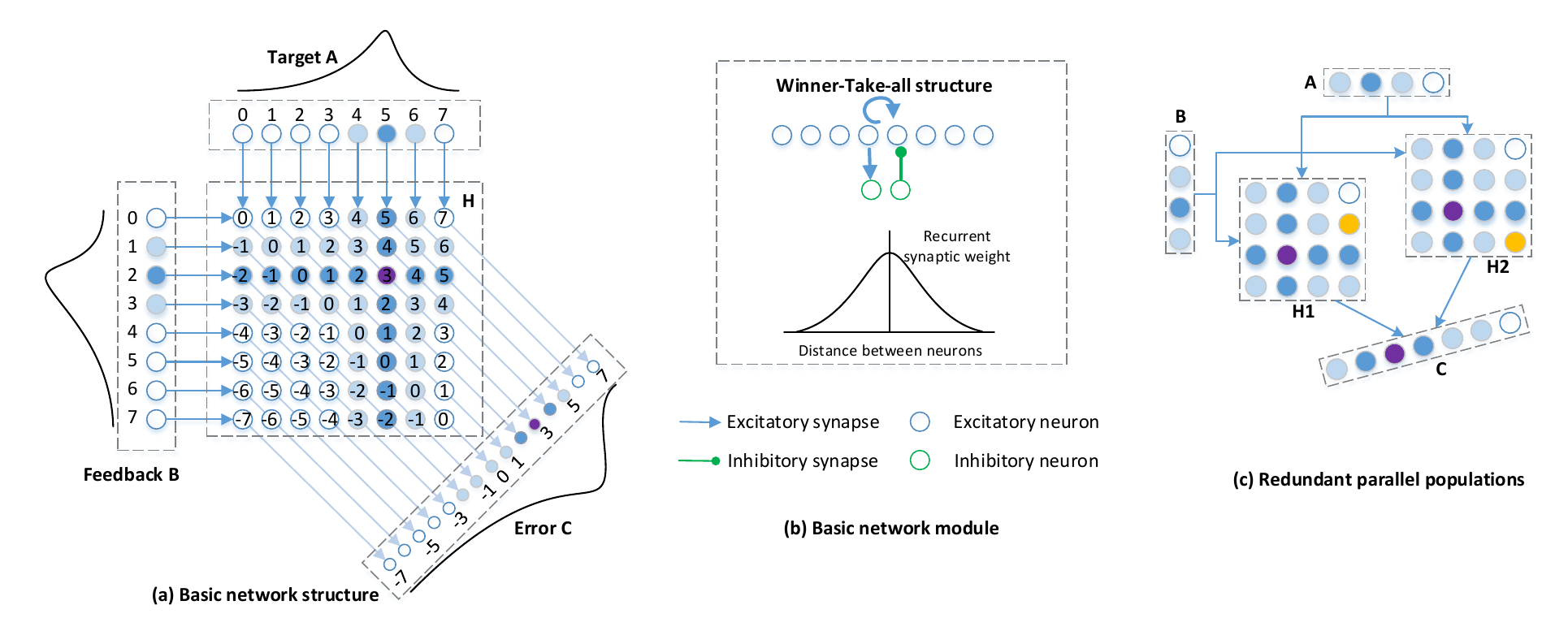}
\caption{(a) Threeway relational network, modified from \cite{zhao2020neuromorphic}. (b) \ac{WTA} implemented in A, B, C and H (in 2D). (c)  Network optimization to further reduce the effect of the mismatch.}
\label{fig:net}
\end{figure*}

Here we are proposing a spiking \ac{P} controller based on a threeway relational network~\cite{diehl2016learning}. 
As shown in \figurename~\ref{fig:net}a, the building blocks of the proposed controller consist of four modules of excitatory and inhibitory neurons: 
\begin{itemize}
    \item[A, B] two input populations encode the desired ($a=\theta^*$) and current ($b=\theta$) iCub head position measured by the encoders, respectively
    \item[H] the hidden population implements the relation $a-b=c$ encoded via the synaptic connections with the input/output populations
    \item[C] the output population encodes the difference between the input variables $c=a-b=\theta^*-\theta$, i.e. the error used in the feedback control loop
\end{itemize}
All populations perform \ac{WTA} computation by means of global inhibitory connections, coupled with local- and self-excitation. Self-excitation is built in the input (A, B), output (C) and hidden (H) populations, while lateral-excitation is used in the hidden (H) and the output (C) population.



Each input population is stimulated by a group of Poisson spike generators via feed-forward one-to-one connections. 
An input variable $a$ is converted into spikes by setting the firing rates of the spike generators according to a Gaussian distribution with the Gaussian mean set to $a$. In \figurename~\ref{fig:net}(a), the firing rates curve of population A represents an 8-level variable $a=5$, which mimics the biological neuron tuning curves~\cite{schoups2001practising}. This encoding scheme allows to robustly encode input variables in the presence of distorted inputs and/or noisy neurons, as, if an input neuron fails to spike, its neighbor neuron will still encode the variable value that is closest to the correct one.

For an input population with $n$ spike generators, the rate of Poisson spike generator $i$ is $rate_{i}$ calculated in Equation \ref{ratei}:
\interdisplaylinepenalty=2500
\begin{align}
\mu &= a\cdot(n-1)  \nonumber \\
rate_{i} &= \exp(-\frac{(i -\mu)^2}{2\sigma^2})\times
rate_{max} \label{ratei} \nonumber \\
i &\in[0,n-1] 
\end{align}
where $a$ is the normalised encoded value, $\mu$ is the mean of the Gaussian calculated using the encoded value, neuron ID $i$ ranges in $[0,n-1]$, $\sigma$ is the Gaussian variance which is adjustable and $rate_{max}$ limits the maximum firing rate of the winner neuron which is closest to $\mu$. The resolution of encoding is $\frac{1}{n-1}$. In our implementation, $n=16$, $rate_{max}=250$~Hz and $\sigma=1$.
The subtraction relation $a-b=c$ is encoded by feed-forward inter-population excitatory connections from A and B to the 2D array of neurons in H, and from H to C, as shown in \figurename~\ref{fig:net}(a). Each neuron in A (B) is connected to one column (row) of neurons in H so that each ``k-diagonal'' (from left to right) represents a unique variable value $a-b$. Neurons located along the same diagonal of H represent the same subtraction result (e.g. $6 - 3 = 4 - 1 =3$). The neurons in each k-diagonal of H are connected to one single neuron in C which represents the result of the subtraction. If the number of possible positions represented by A and B is $n$, the populations in H and C will comprise  $n^2$ and $2n-1$ neurons respectively. In the current implementation $n=16$. 
Recurrent connectivity within each population implements \ac{WTA} competition to speed up the network and sharpen the selectivity, decreasing noise and weak activity.

Each neuron in C represents one possible error value. The instantaneous firing rate of each neuron is transformed into an analog value by means of an exponentially-decaying spiking trace ($E$):
\interdisplaylinepenalty=2500
\begin{align}
E &= E + 1 \text{, whenever a spike is generated} \nonumber \\
E &= E\times\exp(-\frac{t-t_{s}}{\tau} ) \label{trace}
\end{align}
where $t$ is the current time, $t_{s}$ is the time of last spike, $\tau$ is the time constant of the decay trace.
The spiking trace is updated in software periodically ($\sim$ every \SI{60}{\ms}).

Normalised error $c$ can be calculated as the \ac{CoM} using the spiking traces $E$ of all neurons in population C as shown in Equation~\ref{decode}.

\interdisplaylinepenalty=2500
\begin{align}
x_c &= \frac{\sum_{j=1}^{2n-1} rate_j\cdot j}{\sum_{j=1}^{2n-1} rate_j} \nonumber
\label{decode} \\
c &= 2x_c - 1 
\end{align}
where $n$ is the number of neurons in the input population, $rate_j$ is the firing rate of the neuron $j$, and the \ac{CoM} $x_c$ is the normalised value of $c$ with the range of $(-1,1)$. 


\subsubsection{Mismatch minimisation}
\label{subsec:mismatch}

Device mismatch is a characteristic of sub-threshold mixed-signal neuromorphic processors that results in different response transfer functions of identical circuit instantiating in different part of the chip. Therefore, despite sharing the same parameters, different silicon neurons can exhibit slightly different output spiking rate for the same input.

In the proposed network, this can have a significant impact on the hidden and output population, H and C, where it introduces noise in H that propagates to the neurons in C, hindering the selection of the correct error. This drawback can be solved by optimizing the network structure at each level. In the input populations A and B, the \ac{WTA} helps the boost of the activation of the correct winner and to suppress the activation of outliers. The synaptic weights of the self-excitation have a Gaussian profile, as shown in \figurename~\ref{fig:net}(b).
In the hidden population H, three optimization mechanisms are implemented. First, a 2D inhibitory neuron population, H', with the same layout of H is introduced. Each inhibitory neuron h' is excited by the corresponding excitatory neuron h in the same location of the hidden matrix. Reversely, the h' helps h to win the competition by inhibiting other excitatory neurons in the same row, column, and minor diagonal. This selective inhibition prevents outliers from firing more than the expected winner. 

Secondly, both self- and lateral-excitation are constructed inside the hidden excitatory neurons with a connection ratio of $1:3$. 
A lateral-excitation stronger than the self-excitation is used to boost the activation of concurrently active neighbouring neurons with respect to isolated neurons firing because of mismatch. 

Thirdly, we added a redundant parallel population with the same excitatory and inhibitory neuron layout and connections, see \figurename~\ref{fig:net}(c). These two hidden populations independently receive the input from A and B and inject output to C. 
This twin structure weakens the impact of outliers since they are hardly located at the same coordinate in both hidden matrices, whereas the winner neurons, which are strongly stimulated by A and B, are at the same position in two matrices. 
In the output population C, we added lateral-excitation to further consolidate
the winner region domination and to reduce the outlier interference. In total, 1090 neurons are used in this network. 

\subsubsection{Direction neurons}
\label{subsec:dirNeurs}
\begin{figure}[!t]
\centering
\includegraphics[width=0.4\textwidth]{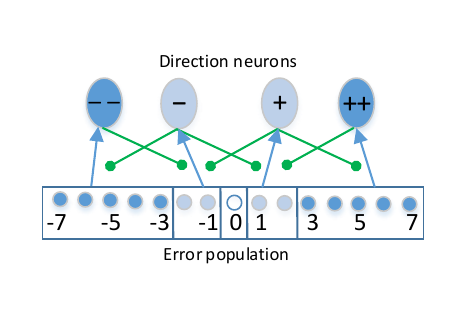}
\caption{Direction neurons to modulate the output command direction.}
\label{fig:dirNeurs}
\end{figure}

As the spiking network encodes the difference between target and encoder position, it is crucial to configure the network parameters such that the population dynamic follows the target input dynamic. Since the threeway network consists of \ac{WTA} populations, the output neurons in C encode an integrated evidence of the feedback error, which results in a trade-off between stability and reaction time to external perturbations. To increase the network sensitivity to the target changes and to mitigate the effect of device mismatch at the level of single neurons, we added another population of inhibitory neurons (hereafter referred to as 'direction neurons'). Specifically, we grouped neurons in C into 4 clusters, according to the encoded error value (\figurename~\ref{fig:dirNeurs}): negative (-), large negative (--), positive (+), large positive (++). Then, we introduced 4 groups of inhibitory neurons receiving excitatory inputs only from one neuron group in C and inhibiting back the other ones. If configured as a hard \ac{WTA}, this connectivity forces competition across neurons sensitive to different error magnitudes (and error sign) and therefore helps the network reacting to fast external perturbations. 

\section{Results}
\label{sec:results}

\subsubsection{Discrete target pursuit task}
\label{subsec:raster}
\begin{figure*}[!t]
\centering
\includegraphics[width=0.7\textwidth]{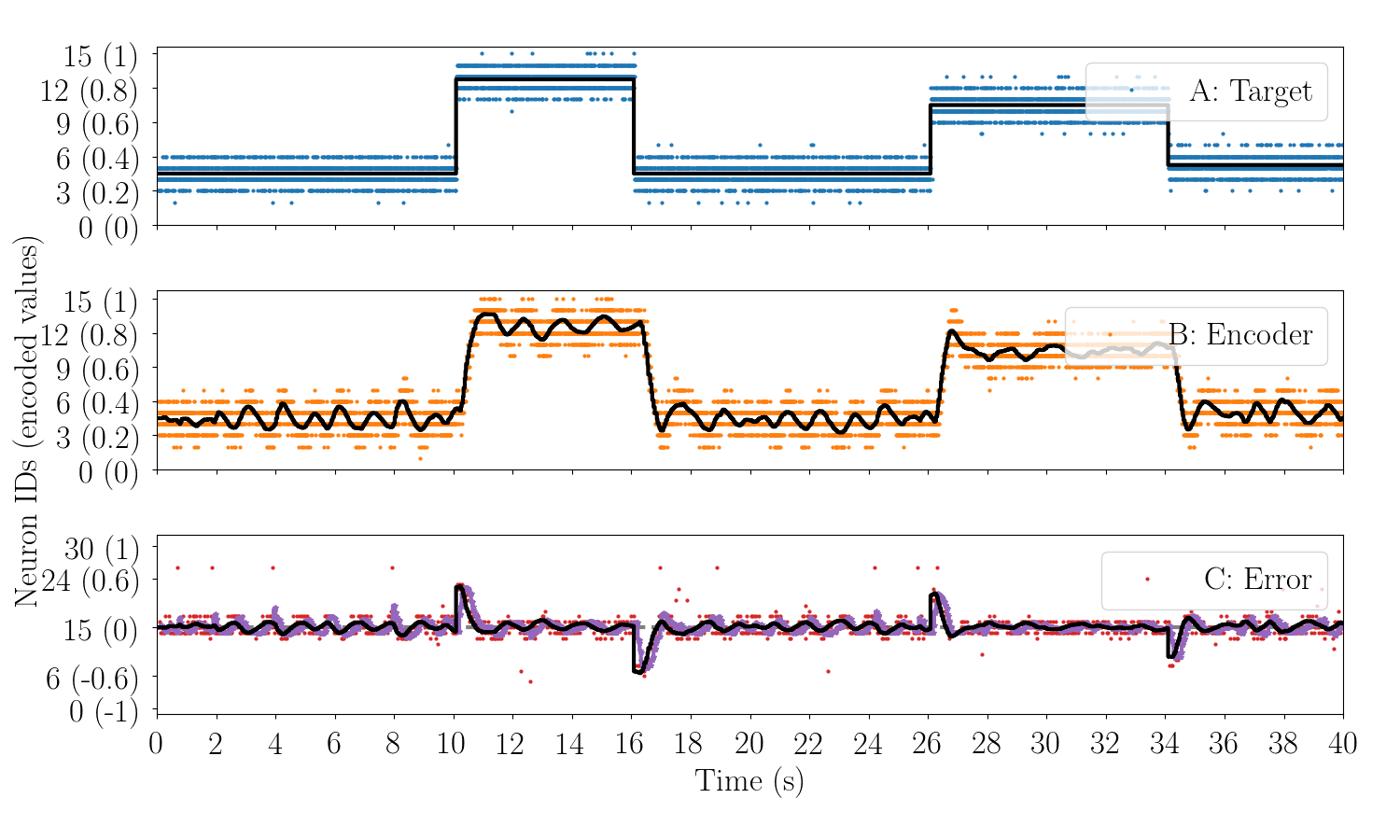}
\caption{Raster plot of input population A, B and output population C during a \ac{DTP} task.}
\label{fig:raster_multi}
\end{figure*}
\figurename~\ref{fig:raster_multi} shows the raster plot of excitatory neurons in populations A, B, and C while the network is fed with the target and encoder positions during a \ac{DTP} task. The black curves in A and B subplots show the input target and encoder positions, while the black one in C subplot is the difference between these two values, i.e., the expected error. The purple curve is the error decoded from the output population C. A, B, and C encode the normalised position $a$, $b$, and $c$. Both $a$ and $b$ range in $[0,1]$, and $c$ in $[-1,1]$. The \ac{CoM} of the firing neurons represents the encoded value of the population. The $y$ axis of \figurename~\ref{fig:raster_multi} shows both the neuron ID and the corresponding encoded value.  $a$ and $b$ are updated in a control loop about every $\SI{20}{\ms}$. 

In the experiment shown in \figurename~\ref{fig:raster_multi}, the starting target position $a$ is set to 0.3 and the control is in steady-state ($b$ = 0.3). Therefore the center of the \ac{WTA} lies in neuron No.4 and No.5, and around neuron No.15 in C. At \SI{10}{\second}, the target position is set to $a=0.85$, resulting in a total error $c=0.55$. Correspondingly, the winner neuron in C moves to No.24. The error generates a positive motor command and actuates the joint to move right towards the new target position. As the controller is closed-loop, the change of the encoder position shifts the \ac{WTA} center in population B between neurons No.8 and No.9. When the encoder reaches the new target position at \SI{10.8}{\second}, the error goes back to zero. At \SI{16}{\second}, the target position is set back to $a=0.3$, which results in a negative error in population C and generates a motor command to move the joint to the left. After the encoder reaches the target, the error returns to zero and the \ac{WTA} center in C moves back to neuron No.15. Each time the controller reaches the steady state, the error stays around zero but still oscillates slightly because of the competition of the neurons around No.15, which leads to the jittery encoder trajectory.

\subsubsection{P controller characterisation}
As with standard \ac{P} controllers, to generate the motor command, the network output error is multiplied by a gain factor $K_p$. The effect of this gain on the controller dynamic was quantified by measuring the \ac{SNN} performance in two scenarios: with a step response task and with the \ac{DTP} task mentioned in Section~\ref{subsec:raster}.

\begin{figure}[!t]
\centering
\includegraphics[width=0.48\textwidth]{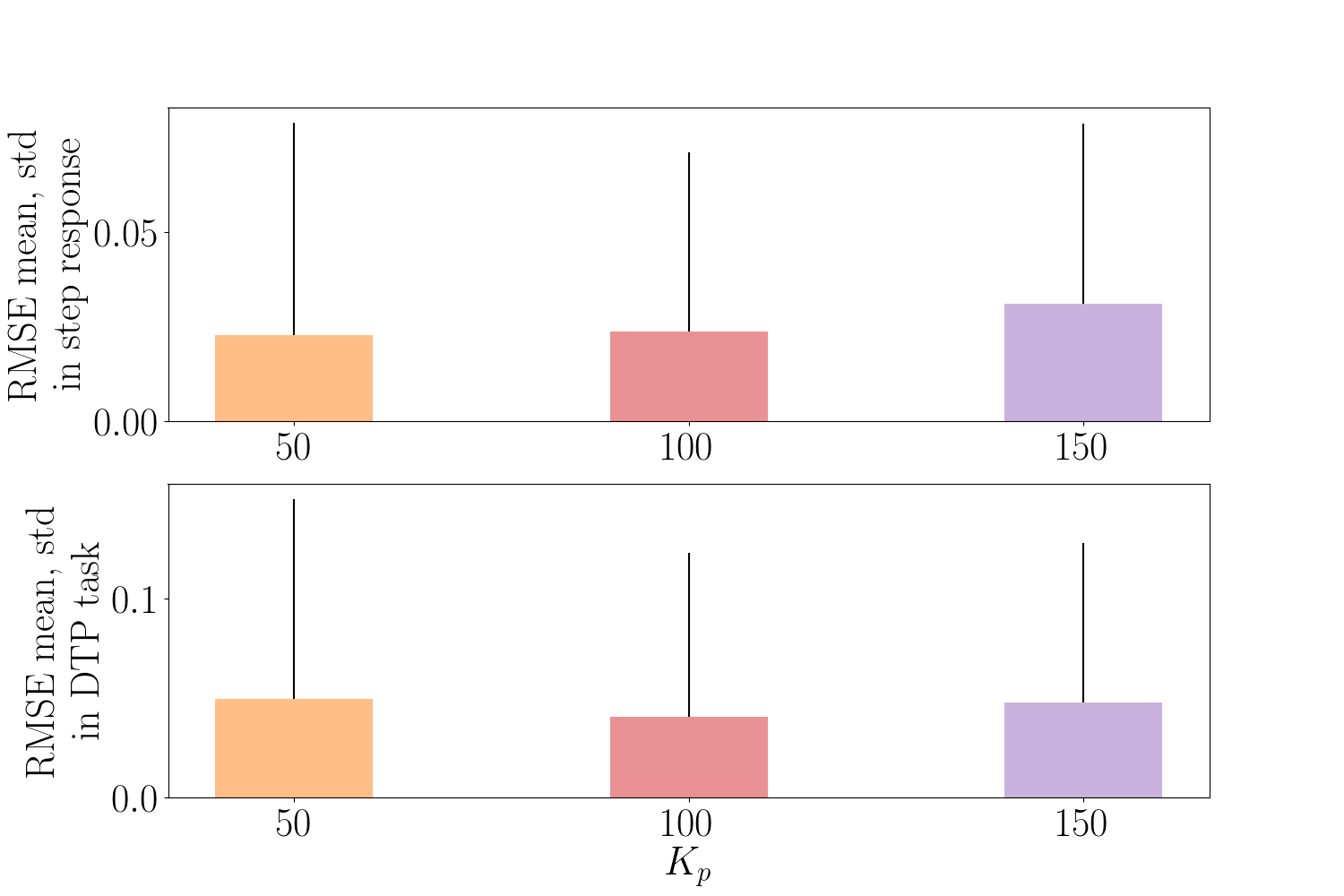}
\caption{\ac{RMSE} with different $K_p$ in the step response and \ac{DTP} task.}
\label{fig:Kp_mse}
\end{figure}

The mean and standard deviation (std) of \ac{RMSE} between target and encoder positions over the first $\SI{40}{\s}$ after the step onset are used to quantify the performance of the controller. \figurename~\ref{fig:Kp_mse} shows the average \ac{RMSE} and its std in the both tasks.


\begin{figure}[!t]
\centering
\includegraphics[width=0.45\textwidth]{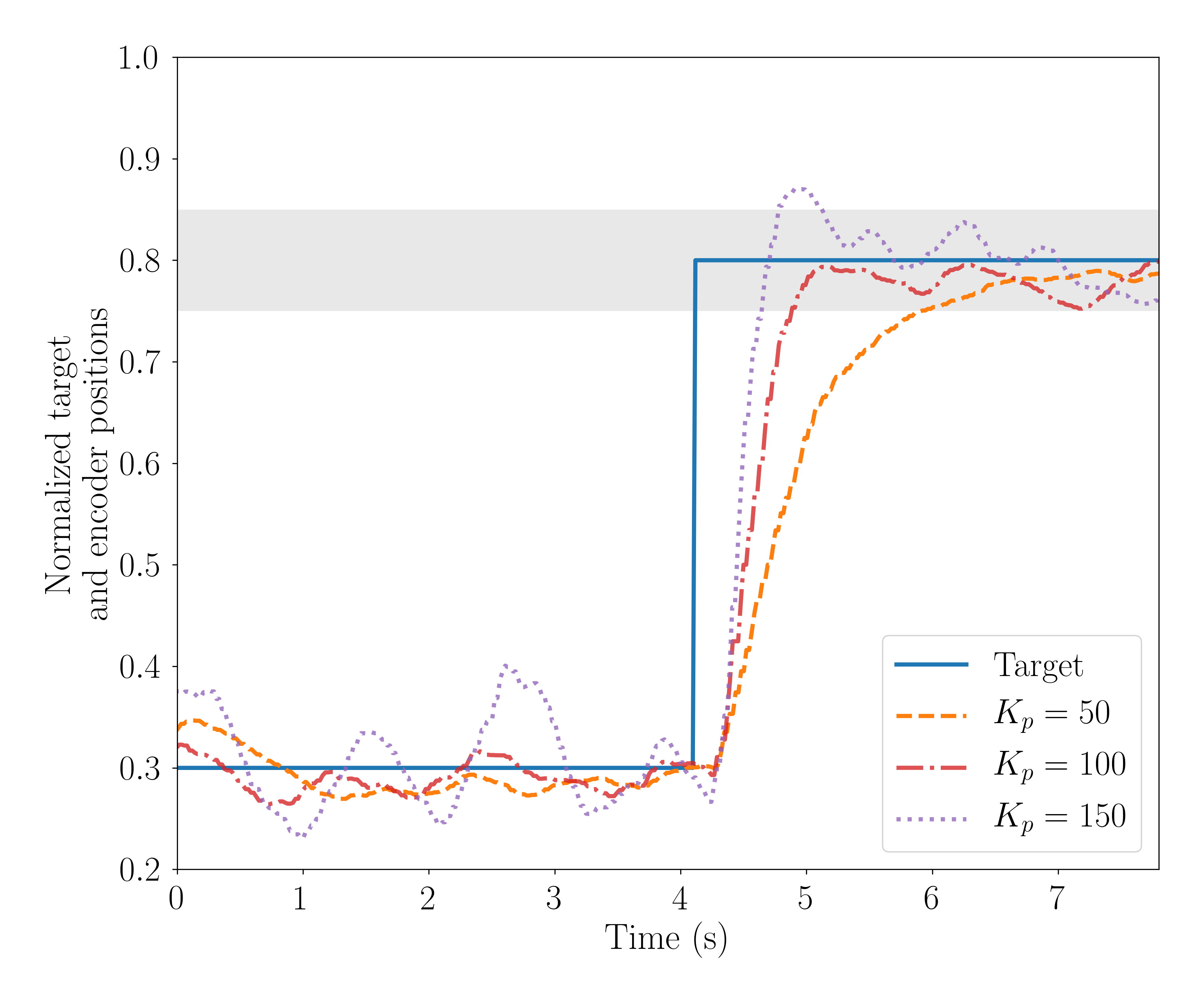}
\caption{Controller step response for three values of $K_p$, with error band as $\pm5\%$ of the target value (grey shading). Average RMSE $= 0.0228, 0.0237, 0.0311$ with $K_p=50,100,150$.}
\label{fig:step}
\end{figure}

\begin{table}[]
\centering
\caption{Rise time with different $K_p$ in the step response}
\scalebox{1.2}{
\begin{tabular}{cccc}
\hline
\multicolumn{1}{|c|}{$K_p$}            & \multicolumn{1}{c|}{50}   & \multicolumn{1}{c|}{100}  & \multicolumn{1}{c|}{150}  \\ \hline
\multicolumn{1}{|c|}{Rise time (s)} & \multicolumn{1}{c|}{1.92} & \multicolumn{1}{c|}{0.89} & \multicolumn{1}{c|}{0.66} \\ \hline
\multicolumn{1}{l}{}                & \multicolumn{1}{l}{}      & \multicolumn{1}{l}{}      & \multicolumn{1}{l}{}     
\end{tabular}}
\label{table:rise_kp}
\end{table}

The performance of the controller with different $K_p$ in the step response task is shown in \figurename~\ref{fig:step}. With the smallest $K_p=50$, the encoder curve is the smoothest, since the controller does not overshoot or oscillate. This results in the smallest \ac{RMSE}, given that the position is updated with small steps, trading off speed with smoothness.
The system rise time, here measured as the time to reach $90\%$ of the target step value, is $\SI{1.92}{\s}$ (see Table~\ref{table:rise_kp}). The encoder trajectory becomes more jittery with larger $K_p$, as the joint position is updated with a larger step. For $K_p=100$, the rise time is $\SI{0.89}{\s}$. With $K_p=150$, the controller reacts faster with rise time of $\SI{0.66}{\s}$ but overshoots the target. After the overshoot, the motor stays around the target position. 

In the \ac{DTP} task, the smallest gain ($K_p=50$), which leads to the slowest controller response, yields the largest average \ac{RMSE} between the target and the encoder curves.
The encoder curve, however, is the smoothest one because deviations from the expected error due to the network noise are less amplified.
However, due to device mismatch, the average \ac{RMSE} does not decrease proportionally to Kp. This is because a larger gain in the control loop can also amplify the network noise if the WTA dynamic is unstable. Thus, larger $K_p$ leads to less smooth encoder curves. 

\begin{figure}[!t]
\centering
\includegraphics[width=0.48\textwidth]{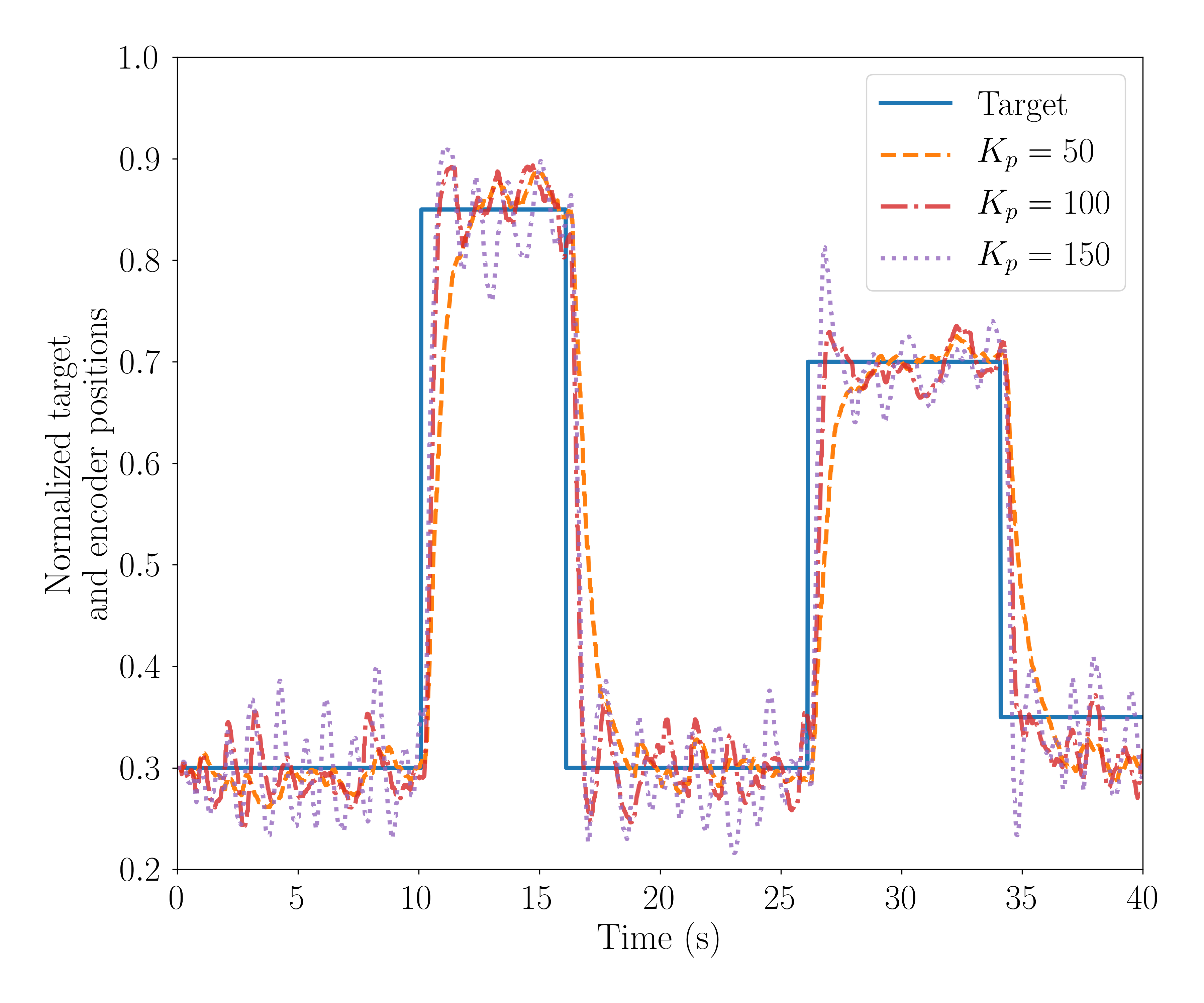}
\caption{The controller performance in the \ac{DTP} task with different $K_p$. Average RMSE $= 0.0494, 0.0407, 0.048$ with $K_p=50,100,150$.}
\label{fig:K_p_fixed}
\end{figure}

\begin{table}[]
\centering
\caption{Rise time with different $\tau$ in the step response}
\scalebox{1.2}{
\begin{tabular}{cccc}
\hline
\multicolumn{1}{|c|}{$\tau$ (s)}       & \multicolumn{1}{c|}{0.005} & \multicolumn{1}{c|}{0.75} & \multicolumn{1}{c|}{5}    \\ \hline
\multicolumn{1}{|c|}{Rise time (s)} & \multicolumn{1}{c|}{0.54}  & \multicolumn{1}{c|}{0.66} & \multicolumn{1}{c|}{0.80} \\ \hline
\multicolumn{1}{l}{}                & \multicolumn{1}{l}{}       & \multicolumn{1}{l}{}      & \multicolumn{1}{l}{}     
\end{tabular}}
\label{table:rise_tau}
\end{table}

\subsubsection{The effect of time constant of the exponentially decaying trace}
\begin{figure}[!t]
\centering
\includegraphics[width=0.48\textwidth]{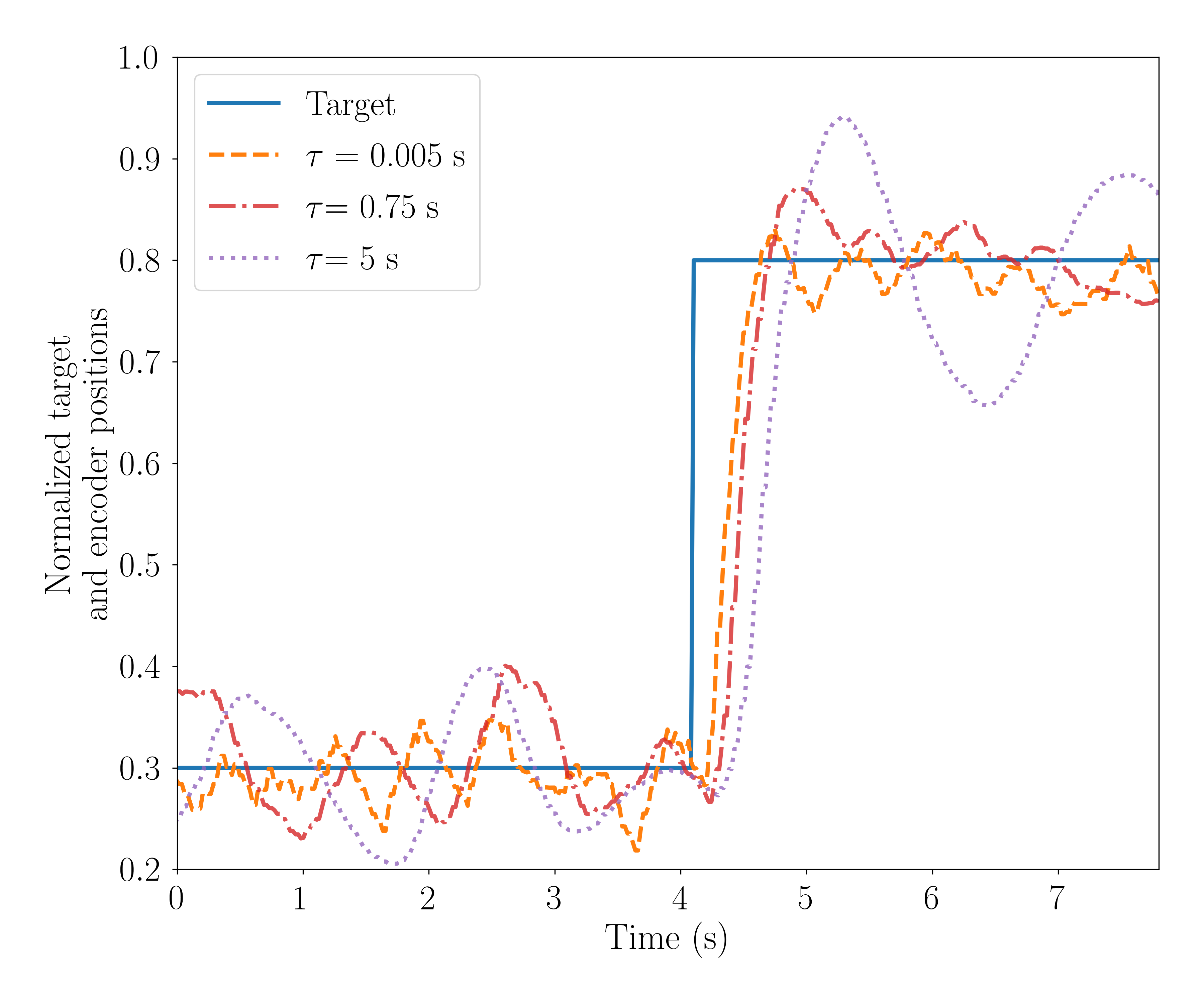}
\caption{The step response of the controller with different $\tau$. Average RMSE $= 0.0254, 0.0311, 0.046$ with $\tau=0.005, 0,75, 5s$.}
\label{fig:traceTau_fixed}
\end{figure}

\begin{figure}[!t]
\centering
\includegraphics[width=0.48\textwidth]{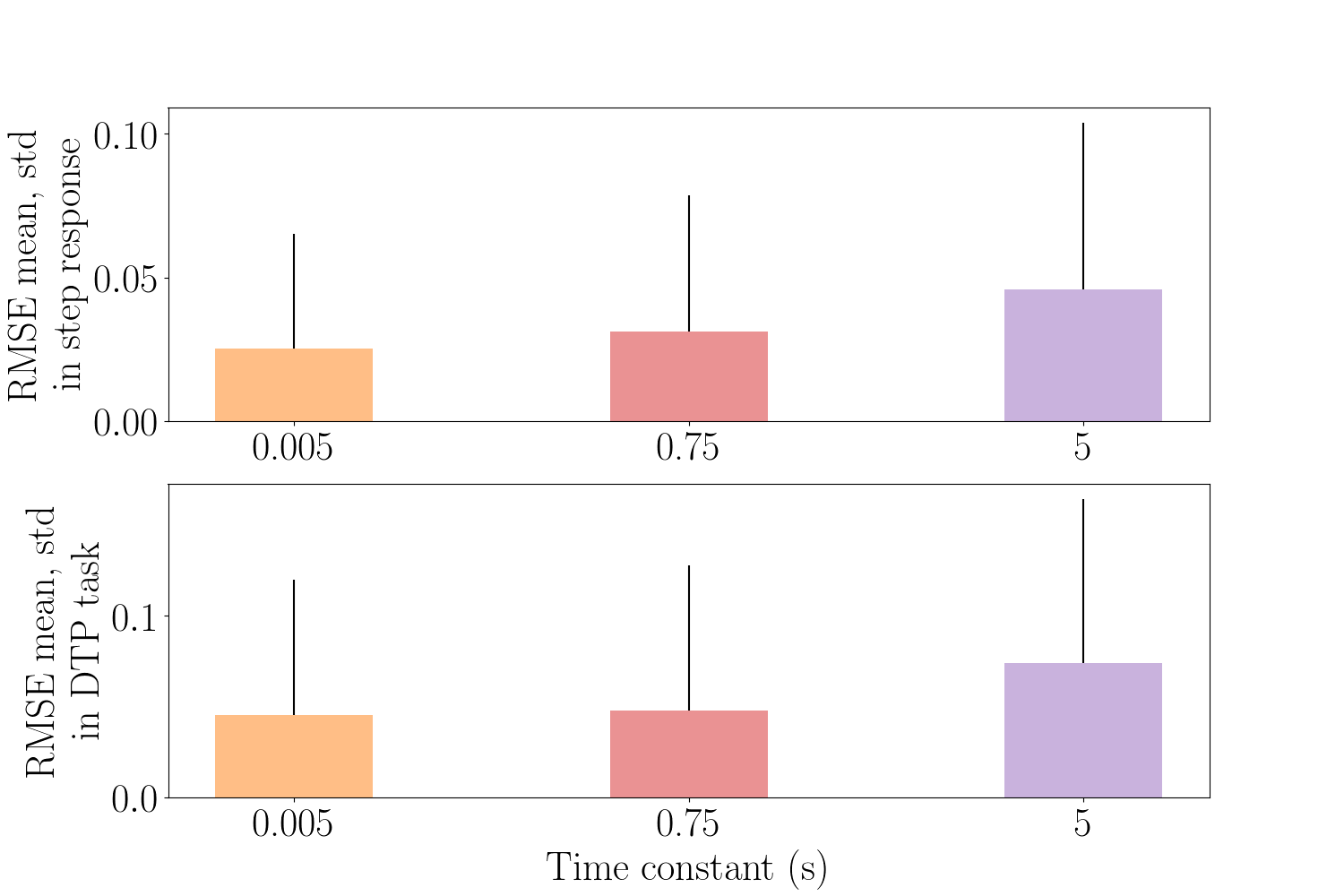}
\caption{\ac{RMSE} with different $\tau$ in the step response and \ac{DTP} task.}
\label{fig:tau_mse}
\end{figure}

One of the key parameters affecting the performance of the spiking controller is the time constant of the exponentially-decaying spiking-trace $E$ (Eq.~\ref{trace}), which tracks the instantaneous firing rate of the output population C. The effect on the controller step response is shown in \figurename~\ref{fig:traceTau_fixed}. The smaller the time constant ($\tau$), the faster the exponential decay and therefore the smaller the integration time window. This speeds up the response of the controller, leading to shorter rise time (see Table~\ref{table:rise_tau}), but increases the control loop sensitivity to the network noise. As a result, the encoder trace jitters around the target position, but the amplitude of the oscillation is relatively small, which leads to a smaller \ac{RMSE}.
The encoder curve gets smoother with larger time constant, but the amplitude and period of oscillation are both larger because the long memory slows down the response of the spiking controller and leads to larger \ac{RMSE}. This trend is quantified in \figurename~\ref{fig:tau_mse} in both the step response and the \ac{DTP} task.

\subsubsection{The controller performance in a sinusoidal target pursuit task}
\begin{figure}[!t]
\centering
\includegraphics[width=0.48\textwidth]{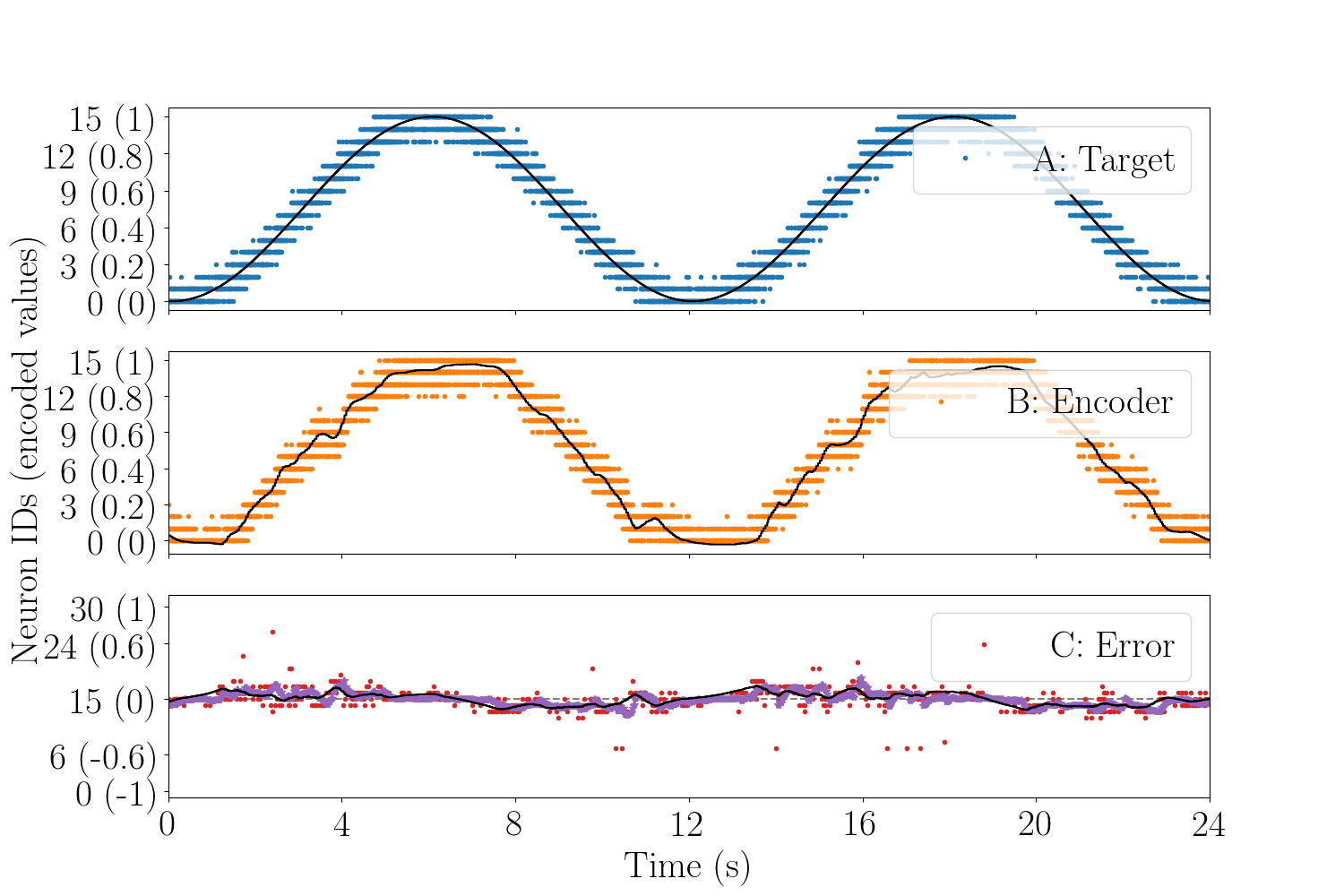}
\caption{The network activity during a sinusoidal target pursuit task.}
\label{fig:raster_sin}
\end{figure}

\begin{figure}[!t]
\centering
\includegraphics[width=0.48\textwidth]{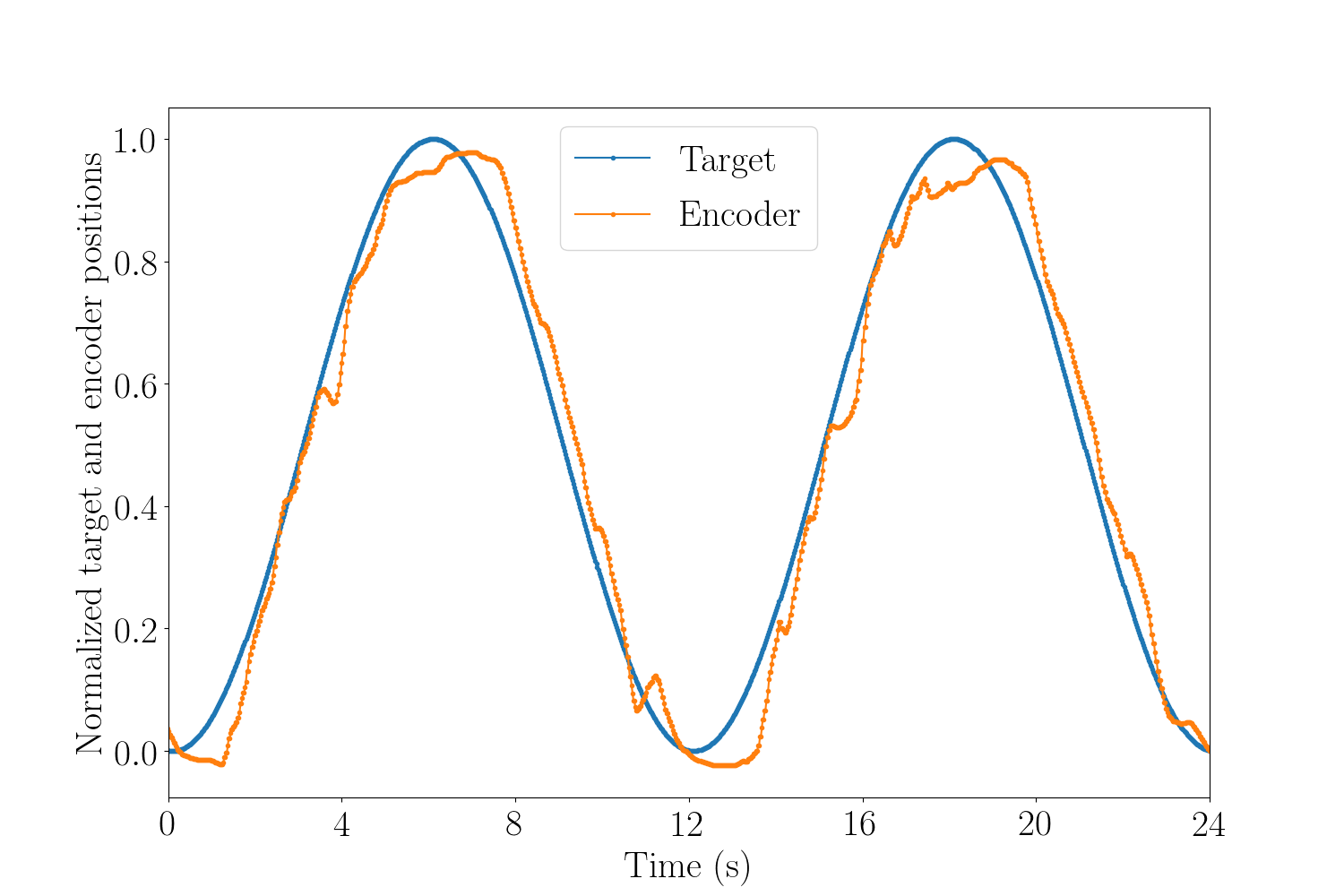}
\caption{The controller behavior during a sinusoidal target pursuit task.}
\label{fig:curve_sin}
\end{figure}

We tested the performance of the neuromorphic controller in a target pursuit task where the target signal is a sinusoidal function with period of $\SI{12}{\s}$. \figurename~\ref{fig:raster_sin} shows the raster plot of excitatory neurons in populations A, B and C during one example trial. As in the DTP experiment, the black curves in subplots A, B and C show the input target position, the input encoder positions and the error between them. The purple curve is the error decoded from the neural activity in C. As the input changes the error encoded by population C follows the expected position error, which leads to the encoder position following the target position (\figurename~\ref{fig:curve_sin}).

\section{Discussion and Conclusion}
\label{sec:discuss}
We presented a spiking P controller implemented on sub-threshold mixed-mode neuromorphic hardware and interfaced with the humanoid robot iCub simulator. We characterised its step response and measured the performance of the neuromorphic controller with a target pursuit task. \figurename \ref{fig:raster_multi} and \ref{fig:raster_sin} show that the controller can react to external perturbations and reach the target state as the distance between the target and encoder position decreases after the initial start-up stage. In this regard, increasing the number of neurons used to encode the positional error would increase the network resolution and therefore the controller performances in tracking the target position. 
As regards the decoding stage, we calculate the \ac{CoM} of the exponentially-decaying spiking trace values of the neurons in the output population in order to find the central location of the firing neurons. The identity of the spiking neuron represents the value of the error, given the space coding paradigm adopted. This value is then sent to the motor using \ac{PWM}. To implement a fully spiking controller, the motor could be moved using \ac{PFM}, where the frequency of the signal encodes for the error value, using rate coding. This could be achieved by an \ac{SNN} structure based on pointer neurons~\cite{hahnloser1999feedback}, which converts space-coding to rate-coding.

The proposed implementation of the closed-loop architecture in a modular system allows easy deployment on the iCub physical robot to test the controller in a real scenario. Validating the \ac{SNN} performances with the model of the iCub robot provides an example framework for comparing different neuromorphic computing platforms with the same control task.

The overall performance of the controller, stability and precision, will be improved adding the Integrative, I, and Derivative, D, terms of the standard \ac{PID}. This could be implemented adding two additional threeway networks, at the cost of drastically increasing the number of required neurons and increasing the latency that hinders the use in real-time and fast applications, as shown recently in RSS20 conference~\cite{rss2020}. To mitigate these drawbacks, we are currently exploring the implementation of the I and D terms exploiting the intrinsic integration and derivative functionalities of synaptic primitives.


\section*{Acknowledgment}
The authors would like to acknowledge the 2019 Capocaccia Neuromorphic Workshop and all its participants for fruitful discussions. 
This work is supported by the Forschungskredit grant FK-18-103, the H2020 ERC project NeuroAgents (Grant No. 724295) and China Scholarship Council. 

\ifCLASSOPTIONcaptionsoff
  \newpage
\fi


\bibliographystyle{IEEEtran}
\bibliography{reference}

\begin{IEEEbiography}[{\includegraphics[width=1in,height=1.25in,clip,keepaspectratio]{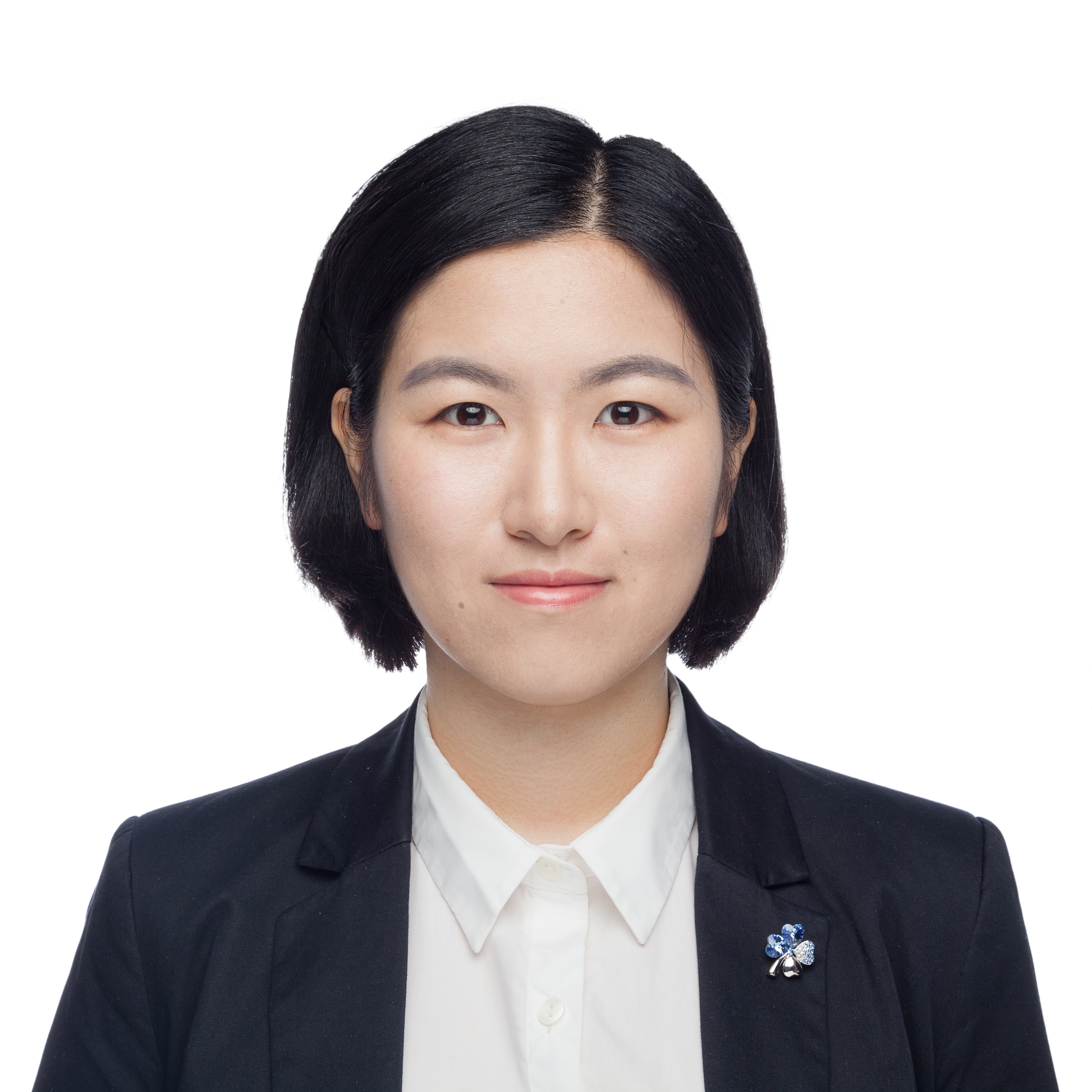}}]{Jingyue Zhao}
received the B.Sc, MSc degree in Computer Science from National University of Defense Technology, Changsha, China in 2015 and 2017, respectively. She studied as a Fellowship PhD student in Computer Science at EPFL, Lausanne, Switzerland in 2017 and switched to the PhD program in Neuroscience at the Institute of Neuroinformatics, University of Zurich and ETH Zurich in December 2018. She is interested in biological motor control circuits and principles, and applying them in robotic applications on neuromorphic hardware.
\end{IEEEbiography}

\begin{IEEEbiography}[{\includegraphics[width=1in,height=1.25in,clip,keepaspectratio]{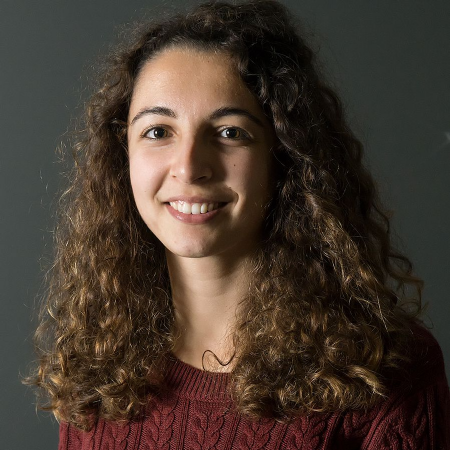}}]{Nicoletta Risi}
studied biomedical engineering at the Universtiy of Genova, Genova, Italy where she graduated (with honors) in 2016. In 2017 she worked as an external collaborator at the bioengineering and robotics research center Centro E. Piaggio, Pisa, Italy. Since September 2017 she is a PhD student at the Institute of Neuroinformatics, University of Zurich and ETH Zurich. She is interested in integrating event-based sensing with spike-based computation on neuromorphic hardware.
\end{IEEEbiography}

\begin{IEEEbiography}[{\includegraphics[width=1in,height=1.25in,clip,keepaspectratio]{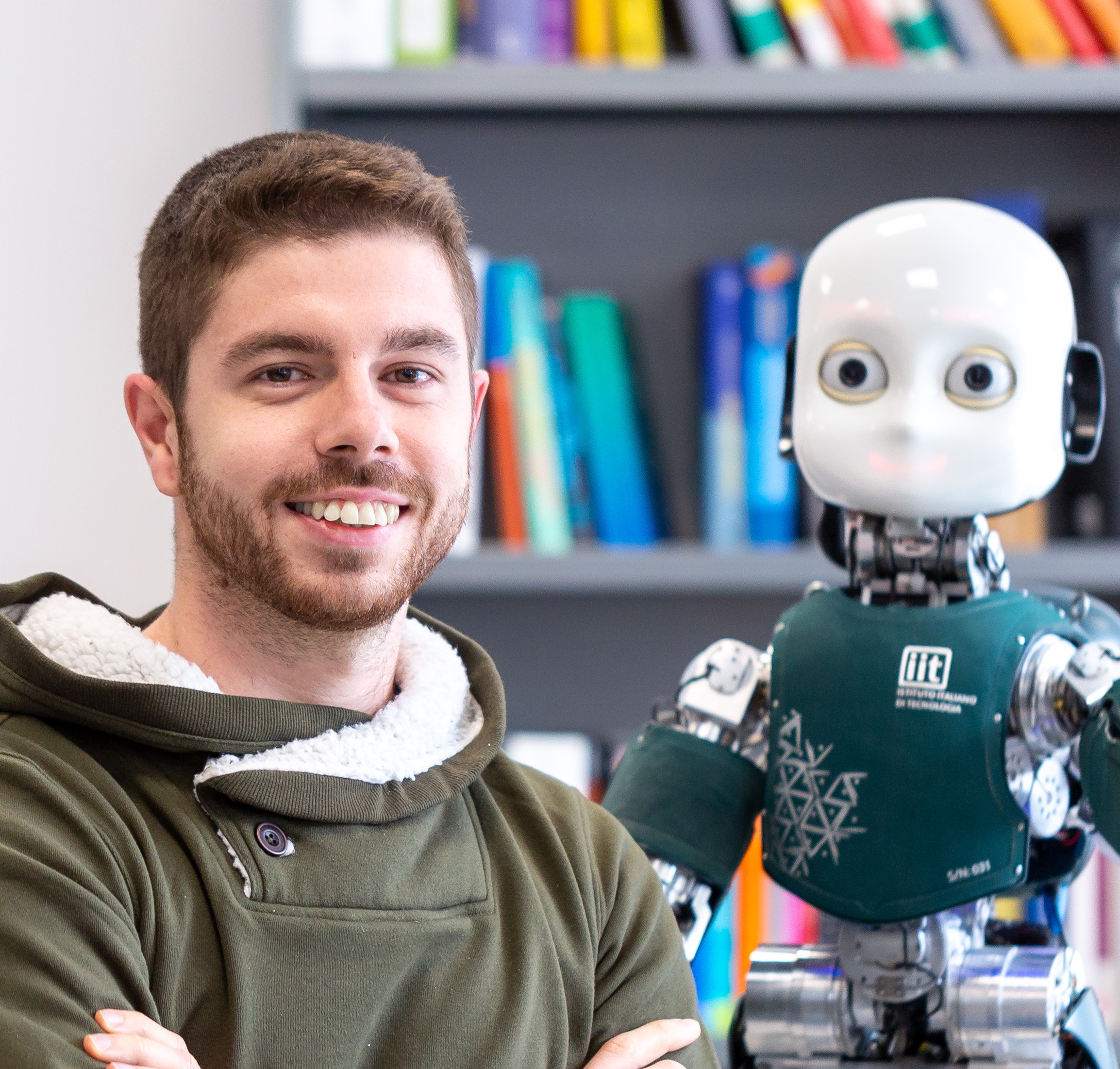}}]{Marco Monforte} is a third year Ph.D. student at Istituto Italiano di Tecnologia, in affiliation with Università degli Studi di Genova (Italy). His doctoral research investigates the use of neuromorphic event-driven cameras for robotics, combining these sensors with deep learning techniques for prediction and robot control in highly dynamic tasks. He received the Laurea degree magna cum laude in Automation Engineering at Università degli Studi di Napoli Federico II, Naples (Italy), in 2017. Other research interests include robot grasping and navigation.
\end{IEEEbiography}

\begin{IEEEbiography}[{\includegraphics[width=1in,height=1.25in,clip,keepaspectratio]{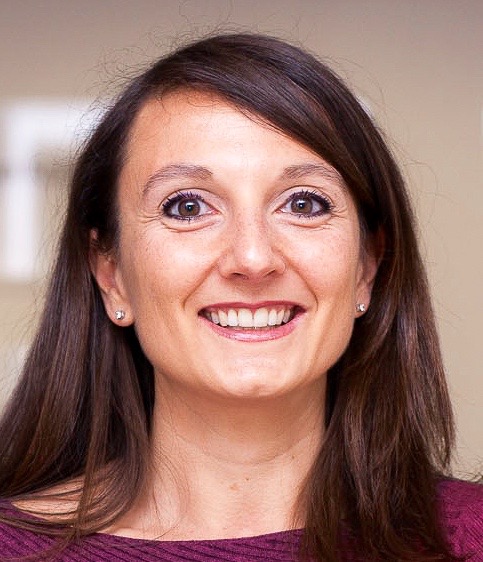}}]{Chiara Bartolozzi}
 (IEEE Member) is researcher at the Istituto Italiano di Tecnologia. She earned a degree in Engineering (with honors) at University of Genova (Italy) and a Ph.D. in Neuroinformatics at ETH Zurich, developing analog subthreshold circuits for emulating biophysical neuronal properties onto silicon and modelling selective attention on hierarchical multi-chip systems. She is currently leading the Event Driven Perception for Robotics group (www.edpr.iit.it), mainly working on the application of the "neuromorphic" engineering approach to the design of sensors and algorithms for robotic perception. She is coordinating the European Training Network NeuTouch. She served as chair of the Neuromorphic Systems and Application Technical Committee of IEEE CAS, currently co-general chair of AICAS2020.
\end{IEEEbiography}

\begin{IEEEbiography}[{\includegraphics[width=1in,height=1.25in,clip,keepaspectratio]{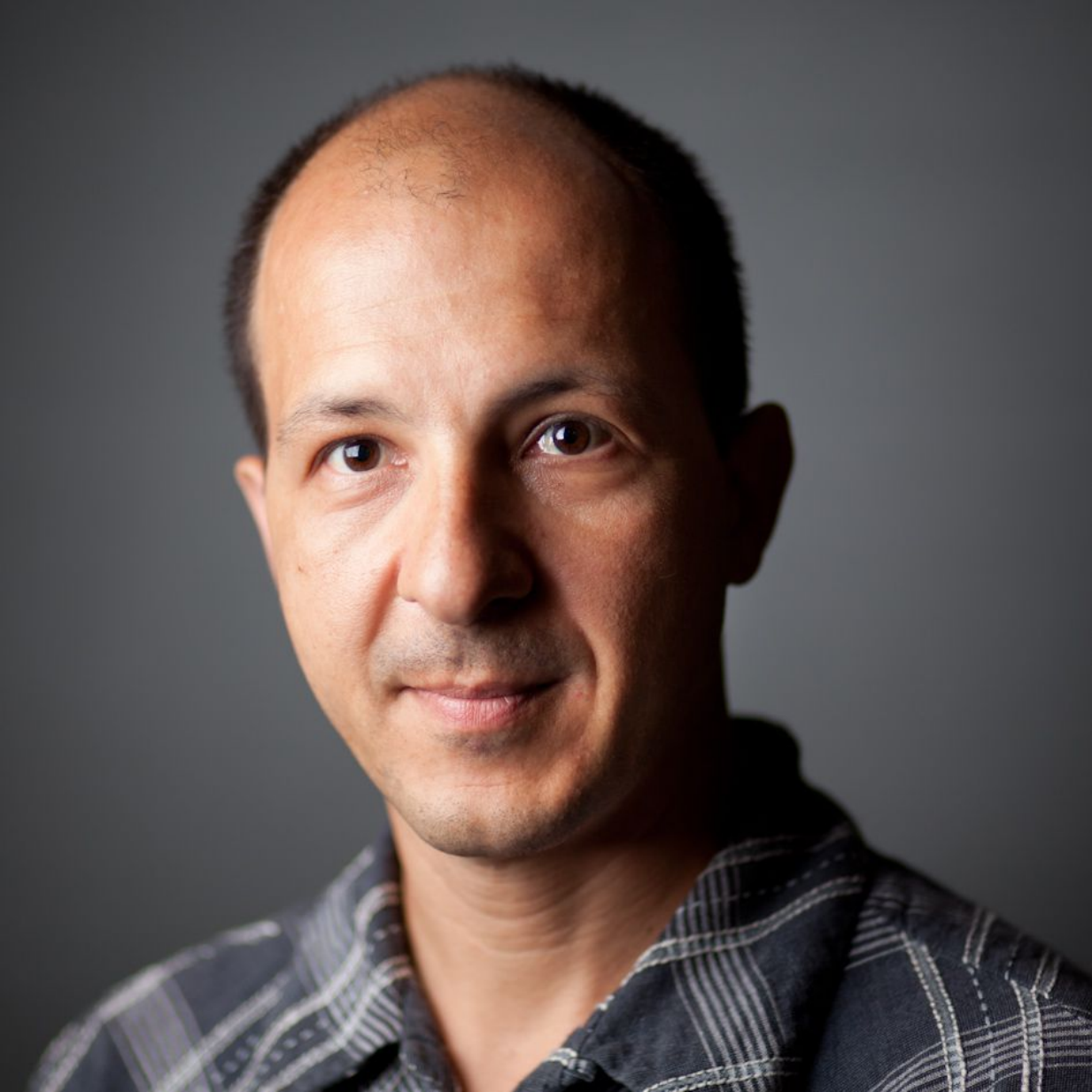}}]{Giacomo Indiveri}
(SM'03) received the M.Sc. degree in electrical engineering and the Ph.D. degree in computer science and electrical engineering from the University of Genova, Genova, Italy, in 1992 and 2004, respectively. Currently, he is a dual Professor at the Faculty of Science of the University of Zurich and at Department of Information Technology and Electrical Engineering of ETH Zurich, Switzerland. He is the director of the Institute of Neuroinformatics (INI) of the University of Zurich and ETH Zurich. He was a post-doctoral research fellow in the Division of Biology at Caltech and at the Institute of Neuroinformatics of the University of Zurich and ETH Zurich. He was awarded an ERC Starting Grant on "Neuromorphic processors" in 2011 and an ERC Consolidator Grant on neuromorphic cognitive agents in 2016.
His research interests lie in the study of real and electronic neural processing systems,
with a particular focus on spike-based learning and spike-based recurrent neural network dynamics. His research and development activities focus on the full custom hardware implementation of real-time sensory-motor systems using analog/digital neuromorphic circuits and emerging memory technologies.

Dr. Indiveri is a member of several technical committees of the IEEE Circuits and Systems Society and a Fellow of the European Research Council.
\end{IEEEbiography}

\begin{IEEEbiography}[{\includegraphics[width=1in,height=1.25in,clip,keepaspectratio]{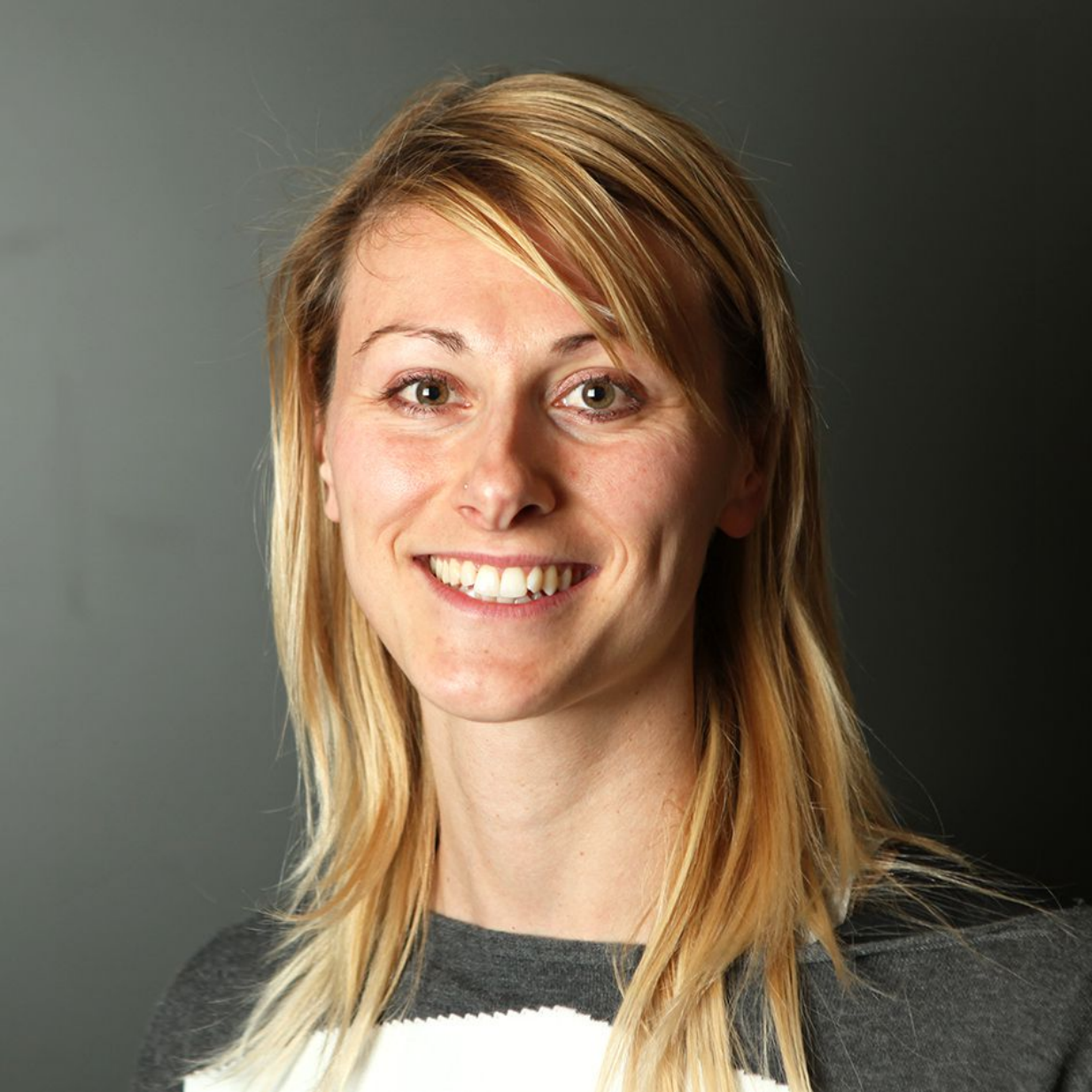}}]{Elisa Donati}
(M'13) received the B.Sc, MSc degree in Biomedical Engineering from University of Pisa, Pisa, Italy (cum laude), and the Ph.D. degree in BioRobotics from the Sant'Anna School of Advanced Studies, Pisa, Italy.

Currently, she is a Research Fellow at the Institute of Neuroinformatics, University of Zurich and ETHZ. Her research activities are at the interface of the neuroscience and the neuromorphic engineering. She is interested in understanding how the biological neural circuits carry out the computation and apply them in biomedical application and neurorobotics. She is investigating how to process biomedical signal to extract features to develop human-robot machine. She is the co-coordinator of the H2020 EU CSA project NEUROTECH.
\end{IEEEbiography}

%








\end{document}